\pdfoutput=1

\documentclass[manuscript,authoryear,11pt]{elsarticle}

\usepackage{tabularx}
\usepackage{graphicx}
\usepackage{amsfonts,amssymb,graphics,epsfig,verbatim,bm,latexsym,amsmath,amsbsy}
\usepackage{amsthm}
\usepackage{longtable}
\usepackage{setspace}
\usepackage[margin=1.2in]{geometry}
\usepackage{csquotes}
\usepackage{caption}

\usepackage[hyphens]{url}
\usepackage{hyperref}

\allowdisplaybreaks
\usepackage{mathtools}
\DeclarePairedDelimiterX{\norm}[1]{\lVert}{\rVert}{#1}
\numberwithin{equation}{section}

\newtheorem{assum}{Assumption}[section]
\newtheorem{thm}{Theorem}[section]
\newtheorem{prop}{Proposition}[section]

\makeatletter
\newcommand{\manuallabelT}[2]{\def\@currentlabel{#2}\label{#1}}
\makeatother

\makeatletter
\newcommand{\manuallabel}[2]{\def\@currentlabel{#2}\label{#1}}
\makeatother

\makeatletter
\newcommand{\manuallabelB}[2]{\def\@currentlabel{#2}\label{#1}}
\makeatother

\makeatletter
\newcommand{\mathleft}{\@fleqntrue\@mathmargin0pt}
\newcommand{\mathcenter}{\@fleqnfalse}
\makeatother

\journal{Journal of Econometrics}

\begin{document}
\sloppy
\begin{frontmatter}



\title{\textbf{Estimating a Large Covariance Matrix in Time-varying Factor Models}}


\author{Jaeheon Jung}
\ead{jj550@economics.rutgers.edu}
\address{Rutgers University, 75 Hamilton St., New Brunswick, NJ 08901, USA}

\begin{abstract}
This paper deals with the time-varying high dimensional covariance matrix estimation. We propose two covariance matrix estimators corresponding with a time-varying approximate factor model and a time-varying approximate characteristic-based factor model, respectively. The models allow the factor loadings, factor covariance matrix, and error covariance matrix to change smoothly over time. We study the rate of convergence of each estimator. Our simulation and empirical study indicate that time-varying covariance matrix estimators generally perform better than time-invariant covariance matrix estimators. Also, if characteristics are available that genuinely explain true loadings, the characteristics can be used to estimate loadings more precisely in finite samples; their helpfulness increases when loadings rapidly change.
\end{abstract}


\end{frontmatter}



\section{Introduction}\label{intro}
A factor model is one of the most widely used methods for estimating large covariance matrices with enhanced precision. By imposing common factor structures on data sets, this model is able to significantly decrease the number of free parameters in the covariance matrix. Many researchers have suggested various types of factor models. \cite{stock2002forecasting}, \cite{bai2002determining}, \cite{bai2003inferential}, and \cite{lam2012factor} study time-invariant factor models, whose factor loadings are fixed over time. However, the assumption that the loadings are fixed for a long period time seems unrealistic because economic transitions, changing technology, and unexpected economic events can change data structures in the long run. Hence, we allow for smooth changes in factor loadings in this article. 

In addition to the above-mentioned structural changes in factor loadings, another assumption imposed on the factor loadings is that the loadings are persistent processes, which implies they are locally constant. However, the assumption can be challenged if the loadings rapidly change (e.g. the financial crisis). In that case, we generally need smaller bandwidth to control a local smoothing bias but the smaller bandwidth inevitably gives rise to the larger variances because of the common bias-variance tradeoff in nonparametric smoothing. To solve the problem, we assume that factor loadings are smooth nonlinear functions of a group of observed characteristics of data sets. This method is based on the idea that, if relevant characteristics are observable, they may help to estimate loadings more accurately.  
 
Under the assumptions mentioned above, this paper proposes two estimators for a high dimensional time-varying covariance matrix. We first estimate a time-varying covariance matrix using a time-varying approximate factor model in which the loadings, factor covariance matrix, and sparse error covariance matrix change smoothly over time. To perform the estimation, we use the local version of principal components analysis (local PCA) introduced by \cite{su2017time} and the principal orthogonal complement thresholding (POET) proposed by \cite{fan2013large}. The other estimator corresponds with a time-varying approximate characteristic-based factor model. We extend the projected principal component analysis (PPCA) proposed by \cite{fan2016projected} to time-varying factor models to perform this estimation. Then, we derive the rates of convergence for each estimated covariance matrix and perform simulation studies to verify the asymptotic results. We also construct global minimum variance portfolios using the estimators and study their out-of-sample performance for practical applications. Our simulation and empirical study indicate that time-varying covariance matrix estimators generally perform better than time-invariant covariance matrix estimators. Also, if characteristics are available that genuinely explained true loadings, the characteristics can be used to estimate loadings more precisely in finite samples. Moreover, their helpfulness increases when loadings rapidly change. 

The following literature is reviewed for this paper. \cite{stock2009forecasting}, \cite{breitung2011testing}, \cite{chen2014detecting}, \cite{han2015tests}, and \cite{cheng2016shrinkage} consider factor models with structural changes in their factor loadings. While these studies focus on accounting for a single radical structural break, \cite{bates2013consistent}, \cite{su2017time}, and \cite{motta2011locally} allow for smooth changes in factor loadings. Corresponding with characteristics-based factor models, \cite{connor2007semiparametric} and \cite{connor2012efficient} assume that factor loadings can be explained entirely by a few observed security characteristics. As one possible framework for estimating characteristics-based factor models, \cite{fan2016projected} propose the PPCA, which applies the conventional PCA to a data matrix projected onto a linear space spanned by covariates. Another type of factor model, the approximate factor model, explores cross-sectional correlations in error covariance matrices. \cite{fan2013large} introduce the POET method to estimate a high dimensional covariance matrix with conditional sparsity structures. 

The rest of this paper is organized as follows. Section \ref{sec2} introduces our models and overviews the sparse covariance matrix estimation. In section \ref{sec3}, we describe two time-varying covariance matrix estimators. In section \ref{sec4}, we state assumptions and establish asymptotic properties for each estimator. Section \ref{sec5} details the implementation of simulation studies. Section \ref{sec6} presents the out-of-sample performance of global minimum variance portfolios. Finally, Section \ref{sec7} concludes. Some assumptions and technical lemmas drawn from \cite{su2017time} are introduced in the \ref{appen}. All proofs are listed in the \ref{suppl}. For the sake of notational simplicity, we use the constant $0<C<\infty$ which varies based on the context. $\lambda_{min}(\bm{A})$ and $\lambda_{max}(\bm{A})$ denote the minimum and maximum eigenvalues of a matrix $\bm{A}$, respectively. The Frobenius norm, spectral norm, and infinity norm are represented by $\lVert \bm{A} \rVert$, $\lVert \bm{A} \rVert _{2}$, and $\lVert \bm{A} \rVert _{\infty}$, respectively.

\setlength{\abovedisplayskip}{6pt}
\setlength{\belowdisplayskip}{6pt}

\section{The Models}\label{sec2}
\subsection{Time-varying Factor Model}\label{subsec2.1}
Consider a time-varying conditional factor model for $N$-dimensional time series with $T$ observations $\{y_{it}\}_{i \leq N, \, t \leq T}:$
\begin{align}
y_{it}=\bm{\lambda}_{it-1}'\bm{f}_{t}+u_{it},   
\label{e.2.1}
\end{align}
where $\bm{f}_{t}=(f_{1t}, \ldots ,f_{Rt})'$ is a $R \times 1$ vector of unobservable common factors, $\bm{\lambda}_{it-1}=(\lambda_{i1t-1}, \ldots ,\lambda_{iRt-1})'$ is a $R \times 1$ vector of corresponding factor loadings, and $u_{it}$ denotes an idiosyncratic error. We assume that both $N$ and $T$ tend to infinity but $R$ is fixed. We model the time-varying $\bm{\lambda}_{it}$ with a function of rescaled time as follows: 
\begin{align*}
\bm{\lambda}_{it}=\bm{\lambda}_{i}(t/T),     
\end{align*}
where $\bm{\lambda}_{i}(\cdot)$ is a nonrandom smooth function on $[0, 1]$. Let $\bm{y}_{t}=(y_{1t}, \ldots ,y_{Nt})'$, $\bm{\Lambda}_{t-1}=(\bm{\lambda}_{1t-1}, \ldots ,\bm{\lambda}_{Nt-1})'$, and $\bm{u}_{t}=(u_{1t}, \ldots ,u_{Nt} )'$. The model (\ref{e.2.1}) can be rewritten in a time indexed vector form:
\begin{align}
\bm{y}_{t}=\bm{\Lambda}_{t-1}\bm{f}_{t}+\bm{u}_{t}. 
\label{e.2.2}
\end{align}
Then, we obtain a conditional covariance matrix $\bm{\Sigma}_{\bm{y}_{t}}=cov(\bm{y}_{t}|\mathcal{F}_{t-1})$:
\begin{align*}
\bm{\Sigma}_{\bm{y}_{t}}=\bm{\Lambda}_{t-1}\bm{\Sigma}_{\bm{f}_{t}}\bm{\Lambda}_{t-1}'+\bm{\Sigma}_{\bm{u}_{t}},  
\end{align*}
where $\mathcal{F}_{t}$ is an information  set up to $t$, and $\bm{\Sigma}_{\bm{f}_{t}}=(\sigma_{ijt}^{f})_{R \times R}$ and $\bm{\Sigma}_{\bm{u}_{t}}=(\sigma_{ijt}^{u})_{N \times N}$ are the covariance matrix of $\bm{f}_{t}$ and $\bm{u}_{t}$, respectively. We allow for smooth changes in both $\sigma_{ijt}^{f}$ and $\sigma_{ijt}^{u}$. Then, similar to $\bm{\lambda}_{it}$,  we model them with functions of rescaled time: 
\begin{align*}
\sigma_{ijt}^{f}=\sigma_{ij}^{f}(t/T)\quad \text{and}\quad \sigma_{ijt}^{u}=\sigma_{ij}^{u}(t/T),   
\end{align*}
where $\sigma_{ij}^{f}(\cdot)$ and $\sigma_{ij}^{u}(\cdot)$ are nonparametric smooth functions on $[0,1]$. Under the assumptions that $\bm{\lambda}_{i}$, $\sigma_{ij}^{f}$, and $\sigma_{ij}^{u}$ are smooth functions, we have the following approximation:\\
For a fixed $r\in \{1,2,\ldots,T\}$,
\begin{align*}
\bm{\Lambda}_{t-1} \approx \bm{\Lambda}_{r-1}, \quad \bm{\Sigma}_{\bm{f}_{t}} \approx \bm{\Sigma}_{\bm{f}_{r}} \quad \text{and} \quad \bm{\Sigma}_{\bm{u}_{t}} \approx \bm{\Sigma}_{\bm{u}_{r}}, \quad \text{when}\,\,\, \frac{t}{T} \approx \frac{r}{T}.  
\end{align*}
It follows that
\begin{align}
\bm{\Sigma}_{\bm{y}_{t}} \approx \bm{\Lambda}_{r-1}\bm{\Sigma}_{\bm{f}_{r}}\bm{\Lambda}_{r-1}'+\bm{\Sigma}_{\bm{u}_{r}}, \quad \text{when}\,\,\, \frac{t}{T} \approx \frac{r}{T}.  
\label{e.2.3}
\end{align}

\subsection{Time-varying Characteristic-based Factor Model}\label{subsec2.2}
\indent It is often the case that factor loadings depend on a group of observed characteristics of data sets. For instance, considering a factor model for stock returns, financial information of each stock such as market capitalization, earnings, and cash flows can affect factor loadings. Characteristic-based factor models capture these kinds of features. 

Let $\bm{X}_{it}=(X_{i1t},\ldots ,X_{idt})'$ be a $d \times 1$ vector of characteristics, where $d$ does not increase. We model factor loadings to be explained entirely by $\bm{X}_{it}$:\footnote{\cite{connor2007semiparametric} also model time-invariant loadings to be explained entirely by characteristics but they assume that each factor loading is a function of each characteristic. \cite{fan2016projected} introduce the generalized time-invariant version of (\ref{e.2.4}), which allows for a component of factor loadings that cannot be explained by characteristics.}
\begin{align}
\bm{\lambda}_{it}=\bm{g}_{t}(\bm{X}_{it}). 
\label{e.2.4}
\end{align}
Here, $\bm{X}_{it}$ changes slowly over time and $\bm{g}_{t}(\cdot)$ is an unknown smooth vector function allowed to be individual-specific and time-varying. So, we in fact assume that in the presence of $\bm{X}_{it}$,  
\begin{align*}
\bm{\lambda}_{i}\left( t/T\right) = \bm{g}_{t}(\bm{X}_{it}). 
\end{align*}
Let $\bm{G}_{t}=(\bm{g}_{t}(\bm{X}_{1t}),\ldots,\bm{g}_{t}(\bm{X}_{Nt}))'$ be a $N \times R$ matrix. Substituting $\bm{\Lambda}_{t-1}$ for $\bm{G}_{t-1}$ in (\ref{e.2.2}) gives a semi-parametric model as follows.
\begin{align}
\bm{y}_{t}=\bm{G}_{t-1}\bm{f}_{t}+\bm{u}_{t}.
\label{e.2.5} 
\end{align}
Then, according to the same procedure illustrated in the previous subsection, we obtain the following approximation of $\bm{\Sigma}_{\bm{y}_{t}}=cov(\bm{y}_{t}| \widetilde{\mathcal{F}}_{t-1})$:
\begin{align}
\bm{\Sigma}_{\bm{y}_{t}}\approx \bm{G}_{r-1}\bm{\Sigma}_{\bm{f}_{r}}\bm{G}_{r-1}'+\bm{\Sigma}_{\bm{u}_{r}}. 
\label{e.2.6}
\end{align}
Note that $\widetilde{\mathcal{F}}_{t}$ denotes a filtration generated by $\{ \bm{\mathsf{X}}_{1},\ldots, \bm{\mathsf{X}}_{t}\}$, where $\bm{\mathsf{X}}_{t}=(\bm{X}_{1t},\ldots,\bm{X}_{Nt})'$.

\subsection{Sparse Matrix}\label{subsec2.3}
In this paper, we assume that $\bm{\Sigma}_{\bm{u}_{t}}$ is a sparse matrix. Following \cite{bickel2008covariance}, the sparsity is measured by the following quantity $m_{t}$:
\begin{align*}
m_{t}=\begin{cases} \max\limits_{i \leq N}\sum_{j=1}^{N}|\sigma_{ijt}^{u}|^{q},& \text{ if } 0<q<1,\\
\max\limits_{i \leq N}\sum_{j=1}^{N}1\{\sigma_{ijt}^{u}\},&\text{ if } q=0,
\end{cases}
\end{align*}
where $1\{\cdot\}$ is the indicator function. We assume that for each fixed $t$, there exists $q\in [0,1)$ to make $m_{t}$ increase slowly and uniformly as $N$ tends to infinity. As mentioned in \cite{fan2013large}, it seems reasonable to assume that an error covariance matrix is sparse in a factor model since after common factors have been taken out, the remaining individual-specific components are likely to be weakly correlated each other. As empirical evidence for the sparse error covariance matrix, \cite{ang2009using} and \cite{ait2017using} observed industry-specific block-diagonal structures in error covariance matrices.

\section{Estimation}\label{sec3}
We propose two time-varying covariance matrix estimators corresponding with the time-varying approximate factor model and the time-varying approximate characteristic-based factor model, respectively.
\subsection{Time-varying Covariance Matrix Estimator without Characteristics}\label{subsec3.1}
We employ the local PCA to estimate each component in the right-hand side of (\ref{e.2.3}). Before describing the estimation, we introduce the following boundary kernel function used for the local PCA:
\begin{align}
k_{h,tr}=h^{-1}K_{h}\left( \frac{t-r}{Th} \right)=\begin{cases} h^{-1}K\left(\frac{t-r}{Th} \right)/\int_{-r/(Th)}^{1}K(u)du,&\text{ if } r\in[1,Th),\\
h^{-1}K\left(\frac{t-r}{Th} \right),&\text{ if } r\in[Th,T-Th],\\
h^{-1}K\left(\frac{t-r}{Th} \right)/\int_{-1}^{(1-r/T)/h}K(u)du,&\text{ if } r\in(T-Th,T],
\end{cases}
\label{e.3.1}
\end{align} 
where $K_{h}(\cdot)$ is a rescaled version of a regular kernel function $K:\mathbb{R} \rightarrow \mathbb{R}^{+}$ and $h=h(T, N)$ is a bandwidth parameter. To make the sum of the kernel function be one in boundaries, we use this boundary kernel function even though it cannot solve the common boundary issue in nonparametric estimation. 

Let $\bm{Y}^{(r)}=(\bm{y}_1^{(r)}, \ldots , \bm{y}_T^{(r)})$ be a $N \times T$ matrix and $\bm{F}^{(r)}=(\bm{f}_1^{(r)}, \ldots , \bm{f}_T^{(r)})'$ be a $T \times R$ matrix, where $\bm{y}_t^{(r)}=(k_{h,tr}^{1/2}\,y_{1t}, \ldots , k_{h,tr}^{1/2}\,y_{Nt})'$ and $\bm{f}_t^{(r)}=(k_{h,tr}^{1/2}\,f_{1t}, \ldots , k_{h,tr}^{1/2}\,f_{Rt})'$. Applying the local PCA to (\ref{e.2.2}), $\widehat{\bm{F}} ^{(r)}=(\widehat{\bm{f}}_1^{(r)}, \ldots , \widehat{\bm{f}}_T^{(r)})'$, the estimated factor matrix, is the $\sqrt{T}$ times eigenvectors corresponding to the $R$ largest eigenvalues of $\bm{Y}^{(r)'}\bm{Y}^{(r)}$ and $\widehat{\bm{\Lambda}}_{r-1}=\bm{Y}^{(r)}\widehat{\bm{F}}^{(r)}/T$ are the estimator of the corresponding factor loadings. Using $\widehat{\bm{F}} ^{(r)}$, we define the estimator of the factor covariance matrix as $\widehat{\bm{\Sigma}}_{\bm{f}_{r}}=(1/T)\sum_{t=1}^{T}\widehat{\bm{f}}_{t}^{(r)}\widehat{\bm{f}}_{t}^{(r)'}$. To obtain the estimator of $\bm{\Sigma}_{\bm{u}_{r}}$, we extend the POET to the time-varying case. Let $\bm{u}_t^{(r)}=(k_{h,tr}^{1/2}\,u_{1t}, \ldots , k_{h,tr}^{1/2}\,u_{Nt})'$ be a $N \times 1$ vector. We use the local residual, $\widehat{\bm{u}}_{t}^{(r)}=\bm{y}_{t}^{(r)}-\widehat{\bm{\Lambda}}_{r-1}\widehat{\bm{f}}_{t}^{(r)}$, for the estimator of $\bm{u}_{t}^{(r)}$ and then calculate $\widehat{\sigma}_{ijr}^{u}=(1/T)\sum_{t=1}^{T}\widehat{u}_{it}^{(r)}\widehat{u}_{jt}^{(r)}$ and $\widehat{\theta}_{ijr}=(1/T)\sum_{t=1}^{T}(\widehat{u}_{it}^{(r)}\widehat{u}_{jt}^{(r)}-\widehat{\sigma}_{ijr}^{u})^{2}$. Based on the rate of convergence of $\widehat{\sigma}_{ijr}^{u}$ and $\widehat{\theta}_{ijr}$, a threshold $\delta_{ijr}$ is defined as follows:\\
\noindent For sufficiently large $C_{NT}>0$,
\begin{align*}
\delta_{ijr}=C_{NT}\,\delta_{NT}\sqrt{\widehat{\theta}_{ijr}}, \quad\quad  \delta_{NT}=\frac{1}{\sqrt{N}}+\sqrt{\frac{\log {NT}}{Th}}+h^{2}\log{T}
\end{align*}
which satisfies
\begin{align*}
\max_{i,j\leq N}\left\vert\frac{1}{T}\sum_{t=1}^{T}\widehat{u}_{it}^{(r)}\widehat{u}_{jt}^{(r)}-E[u_{ir}u_{jr}]\right\vert=O_{p}(\delta_{NT}).
\end{align*}
Compared with the convergence rate of the sample error covariance matrix in \cite{fan2013large}, $\delta_{NT}$ additionally has the third term $h^{2}\log{T}$ and $h$ in the denominator of the second term. $h^{2}\log{T}$ denotes a bias from the time-domain smoothing and the denominator $Th$ means its variance. We discuss the above-mentioned rate in more details in section \ref{subsec4.3}. The thresholding estimator of $\bm{\Sigma}_{\bm{u}_{r}}$ is given by 
\begin{align*}
\widehat{\bm{\Sigma}}_{\bm{u}_{r}}=(\widehat{\sigma}^{u\,\mathcal{T}}_{ijr})_{N\times N},\quad \widehat{\sigma}^{u\,\mathcal{T}}_{ijr}=\begin{cases}1,&\text{if}\,\,i=j,\\
0,&\text{if}\,\left\vert\widehat{\sigma}_{ijr}^{u}\right\vert < \delta_{ijr},\\ 
s_{ij}(\widehat{\sigma}_{ijr}^{u}),&\text{if}\,\left\vert\widehat{\sigma}_{ijr}^{u}\right\vert\ge \delta_{ijr},
\end{cases}
\end{align*}
where $s_{ij}(\cdot)$  is a soft-thresholding function defined as $s_{ij}(z)=sgn(z)(\left\vert z \right\vert - \tau_{ij})_{+}$.\footnote{\cite{fan2013large} verify that various thresholding functions can be used for $s_{ij}(\cdot)$. In this paper, we only use the soft-thresholding function for a simple application.} Plugging in $\widehat{\bm{\Lambda}}_{r-1}$, $\widehat{\bm{\Sigma}}_{\bm{f}_{r}}$, and $\widehat{\bm{\Sigma}}_{\bm{u}_{r}}$ to (\ref{e.2.3}), we finally obtain the estimator of $\bm{\Sigma}_{\bm{y}_{r}}$:
\begin{align*}
\widehat{\bm{\Sigma}}_{\bm{y}_{r}}=\widehat{\bm{\Lambda}}_{r-1}\widehat{\bm{\Sigma}}_{\bm{f}_{r}}\widehat{\bm{\Lambda}}_{r-1}'+ \widehat{\bm{\Sigma}}_{\bm{u}_{r}}. 
\end{align*}
We rigorously study the asymptotic properties of $\widehat{\bm{\Sigma}}_{\bm{y}_{r}}$ and assumptions regarding the estimator in section \ref{subsec4.1}. 

\subsection{Time-varying Covariance Matrix Estimator with Characteristics}\label{subsec3.2}
In the presence of time-varying characteristics of data sets, we extend the PPCA proposed by \cite{fan2016projected} to a time-varying factor model. Specifically, we project local observations onto a linear space spanned by local characteristics and then apply the conventional PCA to the projected local observations. To perform the local PPCA estimation, we employ the sieve method to estimate $\bm{G}_{t}$ nonparametrically. 

Let $\bar{\bm{\phi}}_{t}(\cdot)=(\phi_{1t}(\cdot),\ldots,\phi_{Jt}(\cdot))'$ be a $J \times 1$ vector of basis functions. Here, $J$ denotes the number of sieve terms, and increases slowly as $N$ tends to infinity. The basis functions span a dense linear space of the functional space for $\bm{G}_{t}$. Define a $N\times Jd$ matrix of the basis functions, $\bm{\Phi }_{t}=(\bm{\phi}_{1t},\ldots,\bm{\phi}_{Nt})'$, where $\bm{\phi}_{it}'=(\bar{\bm{\phi}}_{t}(X_{i1t})',\ldots,\bar{\bm{\phi}}_{t}(X_{idt})')$. We essentially assume
\begin{align*}
\bm{G}_{t} \approx \bm{\Phi }_{t}\bm{B}_{t}, 
\end{align*}
where $\bm{B}_{t}$ is a $Jd \times R$ matrix of sieve coefficients. Then, the model (\ref{e.2.5}) can be written as
\begin{align}
\bm{y}_{t}^{(r)}\approx \bm{\Phi }_{r-1}\bm{B}_{r-1}\bm{f}_{t}^{(r)}+\bm{u}_{t}^{(r)}\quad \text{when}\,\,\, \frac{t}{T} \approx \frac{r}{T}.  
\label{e.3.2}
\end{align}
Using (\ref{e.3.2}), we construct the following local weighted least squares problem:\\ 
For $r\in\{1, 2, \ldots , T\}$,
\begin{align}
\min_{\bm{B}_{r-1}, \{\bm{f}_t\}_{t=1}^{T}} \frac{1}{NT}\sum_{i=1}^{N}\sum_{t=1}^{T}[y_{it}^{(r)}-\bm{\phi}_{ir-1}'\bm{B}_{r-1}\bm{f}_{t}^{(r)}]^{2}\quad\quad\quad \label{e.3.3}\\
\text{s. t. }\bm{F}^{(r)'}\bm{F}^{(r)}/T=\bm{I}_{R\times R}, \quad \bm{G}_{r-1}'\bm{G}_{r-1}=\text{diagonal matrix}.\nonumber
\end{align}
Let $\bm{P_{t}}=\bm{\Phi}_{t}(\bm{\Phi}_{t}'\bm{\Phi}_{t})^{-1}\bm{\Phi}_{t}'$ be a $N \times N$ projection matrix. The following Proposition \ref{p.3.1} shows that the solutions of the minimization problem (\ref{e.3.3}) provide the estimators of $\bm{f}_{t}^{(r)}$ and $\bm{G}_{r-1}$.

\begin{prop}\label{p.3.1} Suppose that $\widetilde{\bm{f}}_{1},\ldots,\widetilde{\bm{f}}_{T}$, and $\widetilde{\bm{B}}_{r-1}$ are solutions to (\ref{e.3.3}). Let $\widehat{\bm{F}}^{(r)}=(k_{h,1r}^{1/2}\widetilde{\bm{f}}_{1},\ldots, k_{h,Tr}^{1/2}\widetilde{\bm{f}}_{T})'$ and $\widehat{\bm{G}}_{r-1}=\bm{\Phi}_{r-1}\widetilde{\bm{B}}_{r-1}$. Then, $\widehat{\bm{F}}^{(r)}/\sqrt{T}$ is the eigenvectors corresponding to the first $R$ largest eigenvalues of the $T \times T$ matrix  $\bm{Y}^{(r)'} \bm{P}_{r-1}\bm{Y}^{(r)}$ and $\widehat{\bm{G}}_{r-1}=T^{-1} \bm{P}_{r-1}\bm{Y}^{(r)}\widehat{\bm{F}}^{(r)}$.
\end{prop}
We define $\widehat{\bm{\Sigma}}_{\bm{f}_{r}}$ and $\widehat{\bm{\Sigma}}_{\bm{u}_{r}}$ in the same way as the previous subsection. The only change is the threshold parameter for $\widehat{\bm{\Sigma}}_{\bm{u}_{r}}$ because the convergence rate of the sample error covariance matrix changes when we apply the characteristics to estimating the factor loadings. Let a pre-determined positive decreasing sequence $\omega_{NT}$ to be such that
\begin{align*}
\max_{i,j\leq N}\left\vert\frac{1}{T}\sum_{t=1}^{T}\widehat{u}_{it}^{(r)}\widehat{u}_{jt}^{(r)}-E[u_{ir}u_{jr}]\right\vert=O_{p}(\omega_{NT}). 
\end{align*}
We prove that in the presence of characteristics,
\begin{align*}
\omega_{NT}=\delta_{NT}+ J^{-\eta}.
\end{align*}
Hence, the estimator of $\bm{\Sigma}_{\bm{u}_{r}}$ is given by\\
\begin{align*}
\widehat{\bm{\Sigma}}_{\bm{u}_{r}}=(\widehat{\sigma}^{u\,\mathcal{T}}_{ijr})_{N\times N},\quad \widehat{\sigma}^{u\,\mathcal{T}}_{ijr}=\begin{cases}1,&\text{if}\,\,i=j,\\
0,&\text{if}\,\left\vert\widehat{\sigma}_{ijr}^{u}\right\vert < \omega_{ijr},\\ 
s_{ij}(\widehat{\sigma}_{ijr}^{u}),&\text{if}\,\left\vert\widehat{\sigma}_{ijr}^{u}\right\vert\ge \omega_{ijr},
\end{cases}
\end{align*}\\
where $\omega_{ijr}=C_{NT}\,\omega_{NT}\sqrt{\widehat{\theta}_{ij}^{(r)}}$. Compared with $\delta_{NT}$, $\omega_{NT}$ has the additional term $J^{-\eta}$ related to the rate of convergence of $\widehat{\bm{G}}_{r-1}$. We discuss $\omega_{NT}$ in more detail in section \ref{subsec4.3}. Plugging in $\widehat{\bm{G}}_{r-1}$, $\widehat{\bm{\Sigma}}_{\bm{f}_{r}}$, and $\widehat{\bm{\Sigma}}_{\bm{u}_{r}}$ to (\ref{e.2.6}), we obtain the following substitution estimator of $\bm{\Sigma}_{\bm{y}_{r}}$:
\begin{align*}\\
\widehat{\bm{\Sigma}}_{\bm{y}_{r}}^{P}=\widehat{\bm{G}}_{r-1}\widehat{\bm{\Sigma}}_{\bm{f}_{r}}\widehat{\bm{G}}_{r-1}'+\widehat{\bm{\Sigma}}_{\bm{u}_{r}}. 
\end{align*}
We state assumptions and asymptotic properties for $\widehat{\bm{\Sigma}}_{\bm{y}_{r}}^{P}$ in section \ref{subsec4.2}.

\section{Assumptions and Asymptotic Properties}\label{sec4}
We seperately study the asymptotic properties of two covariance matrix estimators.
\subsection{Time-varying Covariance Matrix Estimator without Characteristics}\label{subsec4.1}
We directly require some assumptions in \cite{su2017time} employing the local PCA to estimate factors, factor loadings, and errors. The assumptions are listed in Appendix \ref{a.A}. In addition to the assumptions, we impose new conditions to estimate a time-varying covariance matrix as follows.\\

\begin{assum}\label{a.4.1} (Factors, Loadings, and Errors)\\ 
(i) $\bm{f}_{t}$ is independent and $E(\bm{f}_{t}\bm{f}_{t}')=\bm{\Sigma}_{\bm{f}_{t}}$  for some positive definite matrix $\bm{\Sigma}_{\bm{f}_{t}}=(\sigma_{ijt}^{f})_{R \times R}$.\\
(ii) $\bm{u}_{t}$ is independent and $\max\limits_{i\leq N,t\leq T}E(u_{it}|\bm{f}_{t})=0$.\\
(iii) $\lambda_{min}(\bm{\Sigma}_{\bm{u}_{t}}) \geq C_{1}$ and $\norm[\big]{\bm{\Sigma}_{\bm{u}_{t}}}\leq C_{2}$.\\
(iv) $r \in [ \lfloor Th \rfloor,T- \lfloor Th \rfloor ]$, where $\lfloor Th \rfloor$ denotes the integer part of $Th$.\\ 
(v) As $(N,T)\rightarrow \infty$, $h\log{T} \rightarrow 0$.\\
(vi) $\sum_{m=-T}^{T}|cov(f_{kt}^{2},f_{kt+m}^{2})||m|^{p}\leq C$ for all $k \leq R$ and $t \leq T$, where $p=0,1,2$.\\
(vii) $\max\limits_{r,t\leq T}E\lVert N^{-1/2}\bm{\Lambda}_{r-1}^{(1)'}\bm{u}_{t}\rVert ^{4} \leq C$, where $\bm{\Lambda}_{r-1}^{(1)}$ denotes the first derivatives of $\bm{\Lambda}_{r-1}$.\\
\end{assum}

\begin{assum}\label{a.4.2} (Smoothness)\\
(i) $\bm{\lambda}_{i}(z)$ is a differentiable function of $z\in[0,1]$, whose first derivative $\bm{\lambda}_{i}^{(1)}(z)$ satisfies $\exists\,C>0$ : $\sup\limits_{z\in[0,1]}\lVert \bm{\lambda}_{i}^{(1)}(z) \rVert<C$ almost surely.\\
(ii) $\sigma_{ij}^{f}(z)$ and  $\sigma_{ij}^{u}(z)$ are differentiable functions of $z\in[0,1]$, and their $k$th derivatives, denoted by $\sigma_{ij}^{f\,(k)}(z)$ and $\sigma_{ij}^{u\,(k)}(z)$, satisfy $\sup\limits_{z\in[0,1]}\max\limits_{i,j\leq N}|\sigma_{ij}^{f\,(k)}(z)|<C_{1}$ and $\sup\limits_{z\in[0,1]}\max\limits_{i,j\leq N}|\sigma_{ij}^{u\,(k)}(z)|<C_{2}$ for $k=1, 2$, respectively.\\
\end{assum}

\begin{assum}\label{a.4.3} (Kernel Function)\\
The kernel function $K:\mathbb{R}\rightarrow \mathbb{R}^{+}$ is a symmetric and continuously differentiable PDF with support [-1,1] such that
\begin{align*}
\int_{-1}^{1}|z|^{m}K(z)^{n}\,dz\leq C \text{ for } m,n=1,2.\\
\end{align*}
\end{assum} 

\begin{assum}\label{a.4.4} (Exponential Tails)\\
There exist positive $\alpha_{1}$, $\alpha_{2}$, $C_{1}$, and $C_{2}$ such that for any $s>0$, $i\leq N$, and $k\leq R$,
\begin{align*}
 P(|u_{it}|>s)\leq exp\{-(s/C_{1})^{\alpha_{1}}\}, \quad P(|f_{kt}|>s)\leq exp\{-(s/C_{2})^{\alpha_{2}}\}.\\
\end{align*}
\end{assum}

Assumption \ref{a.4.1}(i) and \ref{a.4.1}(ii) require $\bm{f}_{t}$ and $\bm{u}_{t}$ to be serially independent but not identically distributed. Also, Assumption \ref{a.4.4} says that their distributions have exponential-type tails. These assumptions help us simplify the proofs of the convergence of $\widehat{\bm{\Sigma}}_{\bm{f}_{r}}$ and $\widehat{\bm{\Sigma}}_{\bm{u}_{r}}$. We leave the case of weakly dependent $\bm{f}_{t}$ and $\bm{u}_{t}$ to future work. Assumption \ref{a.4.1}(iv) requires that $r$ lies in the interior region because our covariance matrix estimators have the common boundary issue in nonparametric estimation even though the boundary kernel (\ref{e.3.1}) is used. Specifically, when $r\in[\lfloor Th \rfloor,T- \lfloor Th \rfloor]$, $\frac{1}{Th}\sum_{t=1}^{T}(\frac{t-r}{Th})K(\frac{t-r}{Th})=o(1)$ and thus $\max\limits_{i,j\leq N}\left\vert\frac{1}{T}\sum_{t=1}^{T}\widehat{u}_{it}^{(r)}\widehat{u}_{jt}^{(r)}-E[u_{ir}u_{jr}]\right\vert=O(h^{2})$. However, when $r\in[1,\lfloor Th \rfloor ]$, $\frac{1}{Th}\sum_{t=1}^{T}(\frac{t-r}{Th})K(\frac{t-r}{Th})=O(1)$, which leads to $\max\limits_{i,j\leq N}\left\vert\frac{1}{T}\sum_{t=1}^{T}\widehat{u}_{it}^{(r)}\widehat{u}_{jt}^{(r)}-E[u_{ir}u_{jr}]\right\vert=O(h)$. In Assumption \ref{a.4.2}, we define the smoothness conditions. Assumption \ref{a.4.1}(iii) makes $\bm{\Sigma}_{\bm{u}_{t}}$ be well conditioned, and Assumption \ref{a.4.1}(v) ensures that the smoothing bias disappears asymptotically. Assumption \ref{a.4.1}(vi) and \ref{a.4.1}(vii) are given to make our proofs easier.

The following Theorem \ref{t.4.1} shows the rate of convergence of $\widehat{\bm{\Sigma}}_{\bm{u}_{r}}$, $\widehat{\bm{\Sigma}}_{\bm{u}_{r}}^{-1}$ and $\widehat{\bm{\Sigma}}_{\bm{y}_{r}}^{-1}$.
\begin{thm}\label{t.4.1}Suppose that Assumption \ref{a.A} and \ref{a.4.1}-\ref{a.4.4} hold. Then, for a fixed $t$ and a sufficiently large C, $\widehat{\bm{\Sigma}}_{\bm{u}_{r}}$ and $\widehat{\bm{\Sigma}}_{\bm{y}_{r}}$ satisfy
\begin{align*}
&\lVert \widehat{\bm{\Sigma}}_{\bm{u}_{r}}-\bm{\Sigma}_{\bm{u}_{r}} \rVert \,=O_{p}\left( m_{t}\,\delta_{NT}^{\,\,1-q}\right),\\
&\lVert \widehat{\bm{\Sigma}}_{\bm{u}_{r}}^{-1}-\bm{\Sigma}_{\bm{u}_{r}}^{-1} \rVert=O_{p}\left( m_{t}\,\delta_{NT}^{\,\,1-q}\right),\\
&\lVert \widehat{\bm{\Sigma}}_{\bm{y}_{r}}^{-1}-\bm{\Sigma}_{\bm{y}_{r}}^{-1} \rVert=O_{p}\left( m_{t}\,\delta_{NT}^{\,\,1-q}\right).
\end{align*}
\end{thm}
\noindent All proofs of Theorem \ref{t.4.1} are contained in our \ref{suppl}.

\subsection{Time-varying Covariance Matrix Estimator with Characteristics}\label{subsec4.2}
In this subsection, we introduce assumptions imposed on the time-varying approximate characteristic-based factor model. The following assumptions are mainly related to the characteristics functions $\bm{G}_{t}$ and the sieve terms.\\

\begin{assum}\label{a.4.5} (Characteristics Function and Sieve Terms) \\
(i) There are positive constants $C_{1}$ and $C_{2}$ such that with probability approaching one, $C_{1}<\lambda_{min}(N^{-1}\bm{G}_{t}'\bm{G}_{t})<\lambda_{max}(N^{-1}\bm{G}_{t}'\bm{G}_{t})<C_{2}$ for all $t$.\\
(ii) $\max\limits_{k\leq R,i\leq N}E[g_{kt}(\bm{X}_{it})^{2}] <\infty$, where $g_{kt}(\bm{X}_{it})$ denotes the $k$th element of $\bm{g}_{t}(\bm{X}_{it})$.\\
(iii) For each $k\leq R$, $i\leq N$, there are nonparametric functions $(g_{k1t}, \ldots ,g_{kdt})$ such that $g_{kt}(\bm{X}_{it})=\sum_{l=1}^{d} g_{klt}(X_{ilt})$.\\
(iv) The sieve coefficient of $\phi_{jt}(X_{ilt})$, denoted by $b_{kjlt}$, satisfies that for $\eta \ge 2$, $\sup\limits_{x\in \mathcal{X}_{lt}}|g_{klt}(x)-\sum_{j=1}^{J}b_{kjlt}\phi_{jt}(x)|^{2}=O(J^{-2\eta})$ as $J \rightarrow \infty$, where $\mathcal{X}_{lt}$ is the support of the $l$th element of $\bm{X}_{it}$.\\
(v) $\max\limits_{t\leq T}\lVert \bm{G}_{t}-\bm{P}_{t}\bm{G}_{t}\rVert_{\infty}=O_{p}(J^{-\eta})$.\\
(vi) There are positive constants $C_{1}$ and $C_{2}$ such that with probability approaching one, $C_{1}<\lambda_{min}(N^{-1}\bm{\Phi}_{t}'\bm{\Phi}_{t})<\lambda_{max}(N^{-1}\bm{\Phi}_{t}'\bm{\Phi}_{t})<C_{2}$ for each $t$.\\
(vii) $\max\limits_{j\leq J, i\leq N, l\leq d, t\leq T}E(\phi_{jt}(X_{ilt})^{2})<\infty$ and $\max\limits_{k\leq R,j\leq J,l\leq d,t\leq T}b_{kjlt}^{2}<\infty$.\\
(viii) $\max\limits_{k\leq R}E[r_{ikt}^{2}]=O(J^{-2\eta})$ and $\max\limits_{i\leq N,k\leq R,t\leq T}|r_{ikt}|=O_{p}(J^{-\eta})$, where $r_{ikt}$ is each element of $\bm{R}_{t}$ for some $(i,k)$.\\
(xi) $\max\limits_{i\leq N}\lVert \bm{\phi}_{it}\rVert \sqrt{Jh^{2}}\ll1$, $\max\limits_{i\leq N}\lVert \bm{\phi}_{it} \rVert \sqrt{J/N}\ll1$, $\max\limits_{i\leq N}\lVert \bm{\phi}_{it} \rVert \sqrt{J/(Th)}\ll1$, and $1 \ll$ $\max\limits_{i\leq N}\lVert \bm{\phi}_{it}\rVert \sqrt{J}$ $\ll \log{T}$.\\ 
\end{assum}

\begin{assum}\label{a.4.6} (Smoothness)\\ 
$\bm{X}_{i}(z)$ and $\bm{g}_{i}(z)$ are differentiable functions of $z\in[0,1]$. Their first derivatives, denoted by $\bm{X}_{i}^{(1)}(z)$ and $\bm{g}_{i}^{(1)}(z)$, satisfy $\sup\limits_{z\in[0,1]}\max\limits_{i\leq N}\lVert \bm{X}_{i}^{(1)}(z) \rVert<C_{1}$ and $\sup\limits_{z\in[0,1]}\max\limits_{i\leq N}\lVert \bm{g}_{i}^{(1)}(z) \rVert<C_{2}$, respectively.\\
\end{assum}

\begin{assum}\label{a.4.7} (Factors and Errors)\\ 
(i) $\sum_{s=-T}^{T}E(|Cov(f_{mt}f_{nt},f_{mt+s}f_{nt+s}|\widetilde{\bm{X}})s^{p}|) \leq C$ for $p=0,1,2$, where $f_{mt}$ denotes the $m$th element of $\bm{f}_{t}$ and $\widetilde{\bm{X}}$ contains all characteristics up to $T$.\\
(ii) $\bm{u}_{t}$ is independent of $\bm{X}_{it}$ and $\max\limits_{t\leq T}\lVert Var(\bm{u}_{t}|\widetilde{\bm{X}}) \rVert<C$.\\
(iii) $\max\limits_{j\leq N, t\leq T}\sum_{i=1}^{N}|E(u_{it}u_{jt})|< C_{1}$, $\max\limits_{t \leq T}\frac{1}{N}\sum_{i=1}^{N}\sum_{j=1}^{N}\sum_{s=1}^{T}|E(u_{it}u_{js})|< C_{2}$, and \\
$\max\limits_{i\leq N}\frac{1}{NT}\sum_{j=1}^{N}\sum_{m=1}^{N}\sum_{t=1}^{T}\sum_{s=1}^{T}|Cov(u_{it}u_{jt},u_{is}u_{ms})|< C_{3}$.\\
\end{assum}

\noindent Assumption \ref{a.4.5}(i)-(viii) are related to the strength of the characteristic-based loadings and the accuracy of the sieve approximation. They are drawn from \cite{fan2016projected} but extended to time-varying factor models. Note that these conditions are imposed for a fixed $t$ in the interior region. So, we achieve the convergences for pointwise $t$. The asymptotic results can be strengthened to be uniform if these assumptions are strengthened uniformly at $t$. Assumption \ref{a.4.5}(xi) is required to restrict the relative rates between $J$ and $(N, Th)$. By Assumption \ref{a.4.6}, both $\bm{X}_{it}$ and  $\bm{g}_{t}(\bm{X}_{it})$ change slowly over time, which implies that $\bm{\Phi}_{t}$, $\bm{P}_{t}$, and $\bm{R}_{t}$ also change slowly over time. Assumption \ref{a.4.7} restricts the dependence for factors and errors. 

The following Theorem \ref{t.4.2} shows the rate of covergence of $\widehat{\bm{\Sigma}}_{\bm{u}_{r}}$, $\widehat{\bm{\Sigma}}_{\bm{u}_{r}}^{-1}$, and $\widehat{\bm{\Sigma}}_{\bm{y}_{r}}^{P\,-1}$. 
\begin{thm}\label{t.4.2} Suppose that Assumption \ref{a.A} and \ref{a.4.1}-\ref{a.4.7} hold. Then, for a fixed $t$ and a sufficiently large C, $\widehat{\bm{\Sigma}}_{\bm{u}_{r}}$ and $\widehat{\bm{\Sigma}}_{\bm{y}_{r}}^{P}$ satisfy
\begin{align*}
&\lVert \widehat{\bm{\Sigma}}_{\bm{u}_{r}}-\bm{\Sigma}_{\bm{u}_{r}} \rVert \,=O_{p}\left( m_{t}\,\omega_{NT}^{\,\,1-q}\right),\\
&\lVert \widehat{\bm{\Sigma}}_{\bm{u}_{r}}^{-1}-\bm{\Sigma}_{\bm{u}_{r}}^{-1} \rVert=O_{p}\left( m_{t}\,\omega_{NT}^{\,\,1-q}\right),\\
&\lVert \widehat{\bm{\Sigma}}_{\bm{y}_{r}}^{P\, -1}-\bm{\Sigma}_{\bm{y}_{r}}^{-1} \rVert=O_{p}\left( m_{t}\,\omega_{NT}^{\,\,1-q}\right).
\end{align*}
\end{thm}
\noindent Theorem \ref{t.4.2} is proved in our \ref{suppl}.

\subsection{Compare the Asymptotic Results of the Two Covariance Matrix Estimators}\label{subsec4.3}
We demonstrate that the rate of convergence of the covariance matrix estimator without characteristics is $m_{t}\delta_{NT}^{1-q}$ and that of the characteristic-based covariance matrix estimator is $m_{t}\omega_{NT}^{1-q}$. We now offer a detailed comparison of the results with and without characteristics, while illustrating the advantage of using characteristics when available. In high dimensional covariance matrix estimation with factor models, the rates of convergence of covariance matrix estimators generally depend on the rates of convergence of error covariance matrix estimators. In this article, $\delta_{NT}$ and $\omega_{NT}$ are the rates of convergence for the error covariance matrix estimators. They correspond with the time-varying factor model without characteristics and with characteristics, respectively. Recall that the rates of convergence are calculated as followed:
\begin{flalign*}
&(\text{Without characteristics}) &&\delta_{NT}=\frac{1}{\sqrt{N}}+\sqrt{\frac{\log {NT}}{Th}}+h^{2}\log{T},&&\\
&(\text{With characteristics}) &&\omega_{NT}=\frac{1}{\sqrt{N}}+\sqrt{\frac{\log{NT}}{Th}}+h^{2}\log{T}+J^{-\eta}.&&
\end{flalign*}

When the two rates are compared, $\omega_{NT}$ does not appear to be faster than $\delta_{NT}$. This result gives rise to a puzzle: the rate of convergence does not seem to improve, even though we apply additional information (characteristics) to the covariance matrix estimation. We now resolve this apparent contradiction by explaining each element of the above-mentioned rates in detail. Generally speaking, the error rate is determined by four inputs: $(i)$ an estimate of unknown factors, $(ii)$ an estimate of $E[u_{it}u_{jt}]$ taken uniformly over $(i,j)$ even if $\bm{u}_{t}$ were known, $(iii)$ smoothing bias for time-varying models, and $(iv)$ an estimate of unknown loadings. For $\delta_{NT}$, the rate of convergence $\textbf{without characteristics}$, these inputs are:
\begin{longtable}{ p{0.03\textwidth}  p{0.93\textwidth} }
$(i)$& $\displaystyle\frac{1}{\sqrt{N}}$: the error of the estimate of the unknown factors, which is optimal even if the loadings were known.\\
$(ii)$& $\displaystyle\sqrt{\frac{\log{NT}}{Th}}$: the uniform rate of estimation for $E[u_{ir}u_{jr}]$ when $\bm{u}_{it}$ is observable. That is $\max\limits_{i,j\leq N}|T^{-1}\sum_{t=1}^{T}u_{it}^{(r)}u_{jt}^{(r)}-E[u_{ir}u_{jr}]|$, and this rate is also optimal.\\
$(iii)$& $h^{2}\log{T}$: the smoothing bias for the time-varying loadings. This is a common term in nonparametric kernel estimation.\\
$(iv)$& $\displaystyle b_{NT}=\frac{1}{\sqrt{Th}}$: the local rate of estimating the factor loadings when the factors are observable. In the time-varying covariance matrix estimation without characteristics, this rate is optimal as the number of local observations is $O(Th)$.
\end{longtable}
\noindent Combining all terms, $(iv)$ is dominated by $(ii)$. Therefore, $\delta_{NT}$ is the final rate.

Now we turn to the rate of convergence $\textbf{with characteristics}$, $\omega_{NT}$. It is also derived from four sources:
\begin{longtable}{ p{0.03\textwidth}  p{0.93\textwidth} }
$(i)$& $\displaystyle\frac{1}{\sqrt{N}}$: the error of the estimate of the unknown factors.\\
$(ii)$& $\displaystyle\sqrt{\frac{\log{NT}}{Th}}$: the rate of estimation for $E[u_{ir}u_{jr}]$.\\
$(iii)$& $h^{2}\log{T}$: the smoothing bias.
\end{longtable}
\noindent All the above three terms are the same as those used to calculate $\delta_{NT}$. These terms do not improve because they are the oracle estimators. In fact, the benefit of knowing characteristics is derived from the estimation of the factor loadings.
\mathleft
\begin{longtable}{ p{0.03\textwidth}  p{0.93\textwidth} }
$(iv)$& $J^{-\eta}+a_{NT}$: the rate of estimation for $\bm{g}_{r-1}(\bm{X}_{ir-1})$ with the observed $\bm{X}_{ir-1}$.
\end{longtable}
\mathcenter
\noindent Here, $a_{NT}=\max\limits_{i\leq N}\lVert \bm{\phi}_{ir-1} \rVert \sqrt{J}(\frac{1}{N}+\frac{1}{Th}+h(\frac{1}{\sqrt{N}}+\frac{1}{\sqrt{Th}}))$. The rate of convergence for estimating $\bm{g}_{r-1}(\bm{X}_{ir-1})$ is justified in the \ref{suppl}. If the basis functions are bounded and $J$ is small relative to $(N,T)$, $a_{NT}$ becomes negligible. Furthermore, $J^{-\eta}$, the sieve approximation error, diminishes very quickly when the selected characteristics genuinely explain true loadings. Hence, under those conditions, $J^{-\eta}+a_{NT}$ converges much faster than $b_{NT}$. This is the benefit of knowing characteristics.

We have revealed that the hidden benefit of using observed characteristics to determine the rate of convergence of estimated loadings when the loadings depend on characteristics. Now, we offer a more detailed explanation of the benefits of using characteristics. The accuracy of the estimation depends on the number of local observations $O(Th)$. In the context of a time-domain smoothing framework, choosing a smaller value for $h$ reduces the smoothing bias but increases variance. While this bias-variance tradeoff always exists, the tradeoff can be mitigated by applying characteristics to estimation for loadings. Specifically, the local PPCA allows us to reduce the smoothing bias with a lower price for the variance (smaller variance) to pay on estimating the loadings. This implies that the improvement can be even more substantial as loadings change more rapidly, which is the case when a smaller $h$ is more desirable. Our simulation study provides a numerical demonstration of this phenomenon in the next section.

\section{Simulation Analysis}\label{sec5}
We use Monte Carlo simulations to examine the finite sample performance of $\widehat{\bm{\Sigma}}_{\bm{y}_{r}}^{-1}$ and $\widehat{\bm{\Sigma}}_{\bm{y}_{r}}^{P\,-1}$. We consider the following two-factor model of $\{y_{it}\}_{i\leq N, t\leq T}$, whose loadings are functions of two characteristics, $x_{it}^{s}$ and $x_{it}^{m}$:
\begin{align*}
y_{it}=g_{1it-1}(x_{it-1}^{s},x_{it-1}^{m})f_{1t}+g_{2it-1}(x_{it-1}^{s},x_{it-1}^{m})f_{2t}+u_{it},
\end{align*}
where
\begin{gather}
\begin{align*}
&g_{1it}(x_{it}^{s},x_{it}^{m})=\alpha_{0t}+\alpha_{1t}x_{it}^{s}+\alpha_{2t}x_{it}^{s\,2}+\alpha_{3t}x_{it}^{m}+\alpha_{4t}x_{it-1}^{m\,2},&\\
&g_{2it}(x_{it}^{s},x_{it}^{m})=\beta_{0t}+\beta_{1t}x_{it}^{s}+\beta_{2t}x_{it}^{s\,2}+\beta_{3t}x_{it}^{s\,3}+\beta_{4t}x_{it}^{m}+\beta_{5t}x_{it}^{m\,2}+\beta_{6t}x_{it}^{m\,3}.&
\end{align*}
\end{gather}

\subsection{Loadings with Small Degree of Time Variations}\label{subsec5.1}
For each simulation,  the following data generation process is performed.
\begin{longtable}{ p{0.01\textwidth}  p{0.949\textwidth} }
$1.$&We generate $\{ \widetilde{g}_{1it} \}_{t \leq 51}$ and $\{ \widetilde{g}_{2it} \}_{t \leq 51}$ from $\widetilde{g}_{1it}= -5 \times 10^{-4}t(t-51-i/30)$ and $\widetilde{g}_{2it}=2 \times 10^{-5}t(t-25+i/30)(t-51-i/30)$. \\ 
$2.$&We generate $\{ \widetilde{f}_{1s} \}_{s \leq 30}$ and $\{ \widetilde{f}_{2s}\}_{s \leq 30}$ from $\widetilde{f}_{1s} = 0.6\widetilde{f}_{1s-1}+e_{1s}$ and $\widetilde{f}_{2s} = 0.3\widetilde{f}_{2s-1}+e_{2s}$, respectively, where $e_{1s}\sim \mathcal{N}(0,0.64)$ and  $e_{2s}\sim \mathcal{N}(0,0.91)$. Then, we calculate a sample covariance matrix using $\{ \widetilde{f}_{1s} \}_{s \leq 30}$ and $\{ \widetilde{f}_{2s}\}_{s \leq 30}$. By repeating this process 51 times, we obtain $\{ \bm{\Sigma}_{\bm{f}_{t}}\}_{t \leq 51}$. \\
$3.$&We first replace the diagonal elements of an $N \times N$ identity matrix with random values from an uniform distribution $\mathcal{U}(0.9,1.2)$. Next, we assign random values from $\mathcal{U}(0.1,0.3)$ to $N$ randomly-selected, off-diagonal elements of the matrix. Then, using the Matlab package nearestSPD, we make the matrix positive definite. We repeat this process 51 times to obtain $\{ \bm{\Sigma}_{\bm{u}_{t}}\}_{t \leq 51}$.\\ 
$4.$&We compute sample means ($\bm{\mu}_{\bm{x}^{s}}$, $\bm{\mu}_{\bm{x}^{m}}$) and sample covariance matrices ($\bm{\Sigma}_{\bm{x}^{s}}$, $\bm{\Sigma}_{\bm{x}^{m}}$) for the size and momentum characteristics introduced in the following section. Then, we generate $\{ \bm{x}_{t}^{s} \}_{t \leq 51}$ and $\{ \bm{x}_{t}^{m} \}_{t \leq 51}$ from $\mathcal{N}(\bm{\mu}_{\bm{x}^{s}}, \bm{\Sigma}_{\bm{x}^{s}})$ and $\mathcal{N}(\bm{\mu}_{\bm{x}^{m}},\bm{\Sigma}_{\bm{x}^{m}})$, where $\bm{x}_{t}^{s}=(x_{1t}^{s},\ldots,x_{Nt}^{s})'$ and $\bm{x}_{t}^{m}=(x_{1t}^{m},\ldots,x_{Nt}^{m})'$.\\
$5.$&We interpolate all pre-generated data up to a sample size T using the cubic spline interpolation in Matlab. \\
$6.$&We fit $g_{1it}(x_{it}^{s},x_{it}^{m})$ and $g_{2it}(x_{it}^{s},x_{it}^{m})$ to $\widetilde{g}_{1it}$ and $\widetilde{g}_{2it}$, respectively. The fitted functions are treated as true time-varying loading functions.\\
$7.$&$\{ \bm{f}_{t} \}_{t \leq T}$ and $\{ \bm{u}_{t} \}_{t \leq T}$ are generated from $\mathcal{N}(0,\bm{\Sigma}_{\bm{f}_{t}})$ and $\mathcal{N}(0,\bm{\Sigma}_{\bm{u}_{t}})$, respectively.
\end{longtable}
\noindent Note that we are able to obtain simulation data sets closer to the smoothness assumptions by using interpolation, which refines the generated data in local windows.

We generate data sets using a different combination of $T$ and $N$ and estimated $\bm{\Lambda}_{t-1}$ and $\bm{\Sigma}_{\bm{y}_{t}}^{-1}$ using the local PCA and PPCA. The simulation is repeated 500 times. Then, we measure average estimation errors for both $\bm{\Lambda}_{t-1}$ and $\bm{\Sigma}_{\bm{y}_{t}}^{-1}$, applying the Frobenius norm. In this simulation, the number of factors is assumed to be known. For nonparametric estimation, we use the Epanechnikov kernel and select values of$h$ and $C_{NT}$ that satisfy the terms of the following minimization problem: \\
For each $t$,
\begin{align*}
\min_{h\in[0.05,0.3],\,C_{NT}\in[0.1,1.2]}\lVert \widehat{\bm{\Sigma}}_{\bm{y}_{t}}^{-1}-\bm{\Sigma}_{\bm{y}_{t}}^{-1}\rVert.
\end{align*}
We apply polynomial basis functions with the sieve dimension $J=4$ for the local PPCA. 
\begin{figure}[!h]
\centering
\includegraphics[scale=1.0]{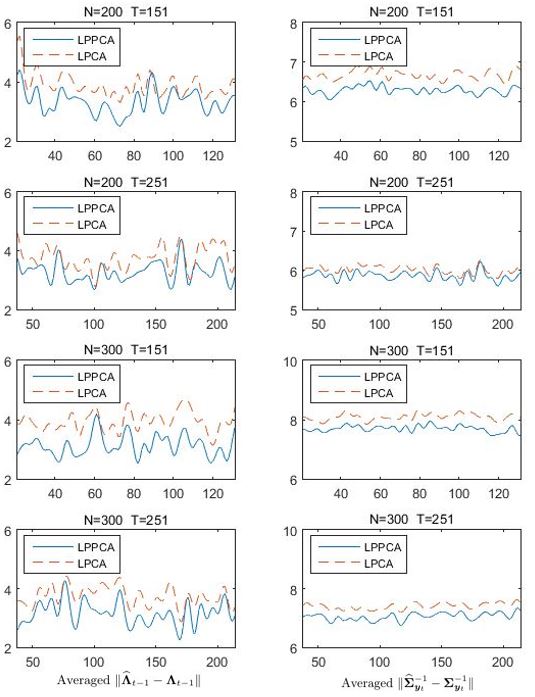}
\caption{Dashed red curves and solid blue curves correspond to the local PCA and the local PPCA, respectively. For $T=151$ and $T=251$ with $N$ values varying between $N=200$ and $N=300$, the left column plots had an average of $\lVert \widehat{\bm{\Lambda}}_{t-1}-\bm{\Lambda}_{t-1}\rVert$ over 500 simulations and the right column plots had an average of $\lVert \widehat{\bm{\Sigma}}_{\bm{y}_{t}}^{-1}-\bm{\Sigma}_{\bm{y}_{t}}^{-1}\rVert$. }\label{fig1}
\end{figure}
\noindent \indent Figure \ref{fig1} displays the simulation results for $T=151$ and $T=251$ with $N$ values varying between 200 and 300. We report only these four cases in order to save space, as other combinations produced similar results to those reported here. The dashed red curve and solid blue curve in the figure denote the local PCA and the local PPCA, respectively. Upon examining the simulation results, we first observe that the local PPCA outperforms the local PCA at all $t$ for both loadings and inverse covariance matrix estimation. Recall that the only difference between the two estimators is whether the observed characteristics are applied to the loading estimation or not. Therefore, we can state that the benefit of considering characteristics in the loading estimation is more precise inverse covariance matrix estimation. The result also supports our discussion in Section \ref{subsec4.3}, in which we state that the benefit of estimating loadings with observed characteristics can be substantial in finite samples. 

Figure \ref{fig1} also illustrates that, as $T$ decreases, the difference between the average estimation errors for the local PPCA and the local PCA increases, given a fixed $N$. This result reinforces the argument made Section \ref{subsec4.3} regarding the degree of variation in true loadings. Note that for both local PCA and local PPCA, $h=0.1$ was chosen at almost all $t$ in this simulation. This implies that the local window size is almost fixed. Also, recall that we interpolated pre-generated data to create the data sets. Thus, given a fixed window size, local data becomes rougher (meaning that there is greater variances) when $T$ is set to be a small number. This means that the benefit of the local PPCA, namely offsetting the bias-variance tradeoff, increases as $T$ decreases. This observation is the main subject of this article. Therefore, we reexamine this result in the following subsection, using different method to generate loadings. 

\subsection{Loadings with a High Degree of Variation}\label{subsec5.2}
\begin{figure}[hbt!]
\centering
\includegraphics[scale=0.68]{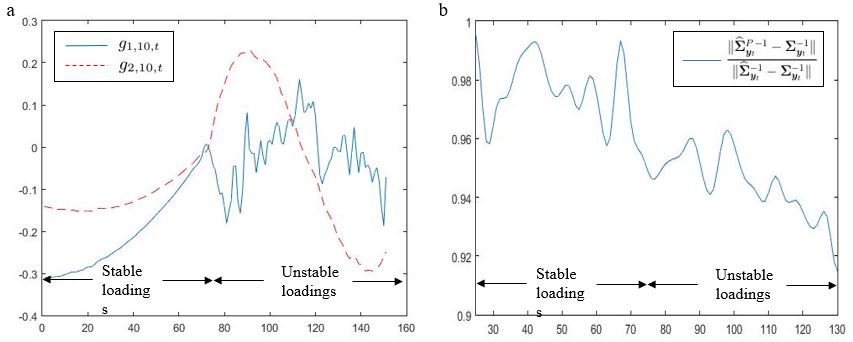}
\caption{(a) True loading curves for $i=10$: $g_{1,10,t}$(solid blue) and $g_{2,10,t}$(dashed red). (b) The ratio of the average estimation errors of the inverse covariance matrix generated by the local PPCA compared to those generated by the local PCA.}\label{fig2}
\end{figure}

To verify the benefit of a local PPCA when factor loadings fluctuate violently, we make a change in the degree of variation of the true loadings. Specifically, we generate $\widetilde{g}_{1,it}=5\times 10^{-4} (t+20+i/30)(t-25)$ and $\widetilde{g}_{2,it}=5 \times 10^{-6} (t-25)(t+15+i/30)(t+50+i/30)$ from t=1 to t=25, while we generate $\widetilde{g}_{1,it}=2\cos(4\pi t/T+ i)$ and $\widetilde{g}_{2,it}=2 \times 10^{-4} (t-25)(t-34-i/30)(t-55)$ from t=26 to t=51. The structural break makes the true loadings change more rapidly in the second half of the sample period. To illustrate the change, we plot the true loading curves of $i=10$ in Figure \ref{fig2}(a). The other data are generated in the same way, and all data are interpolated up to $T=151$. 

We calculate a ratio compareing the average estimation errors of the inverse covariance matrix generated by the local PPCA with those generated by the local PCA and plot the ratio in Figure \ref{fig2}(b). Figure \ref{fig2}(b) illustrates that the ratio is less than one for all values of $t$ and gradually drops after the first half of the sample period. This result indicates that the local PPCA generally performs better the local PCA, and the outperformance becomes more marked in the second half of the sample period. This occurs because, if true loadings are volatile, then the smoothness assumptions imposed on the loadings in our models are not satisfied. It follows that the estimated loadings exhibit a larger bias. As a result, the local PPCA estimator does not work as we expected. Nonetheless, we can compensate for this problem by using the local PPCA. Specifically, by projecting the data onto genuine characteristics, we can make it smoother without increasing bias. The benefit obtained from data smoothing becomes greater as true loadings change more rapidly.

The primary simulation results are summarized as follows.
\begin{longtable}{ p{0.01\textwidth}  p{0.949\textwidth} }
$1.$& Estimating loadings using observed characteristics helps to make estimates of inverse covariance matrices more accurate. This provides a substantial benefit in finite samples.\\
$2.$& The benefit of using characteristics in loading estimation increases when loadings change rapidly.
\end{longtable}

\section{Empirical Analysis}\label{sec6}
In this section, to empirically examine the performance of the local PCA and the local PPCA, we construct global minimum portfolios using various covariance matrix estimators. We then compare their out-of-sample performance.

\subsection{Data and Methodology}\label{subsec6.1}
We use weekly data from 370 stocks, which are randomly selected from all common domestic stocks traded on the NYSE and the AMEX that are not missing values in stock returns, book value of equity, and market value of equity. The data select ranged from January 1998 to December 2016. The stock returns are measured in excess of the risk-free rate derived from the one-month Treasury Bill returns. We collect the data on stock returns and one-month Treasury Bill returns from the Center for Research in Security Prices (CRSP) database, and the book value of equity and the market value of equity from the Compustat database. We also download the Fama-French three factors from the website \url{http://mba.tuck.dartmouth.edu/pages/faculty/ken.french/data\_library.html}. Four characteristics of each stock used for characteristic-based factor estimators (size, value, momentum, and volatility) are derived following the guideline established by \cite{connor2012efficient}. 

We construct global minimum variance portfolios using various covariance matrix estimators and evaluate their out-of-sample performance following the methodology described by \cite{chan1999portfolio}. Specifically, we estimate the covariance matrix for the returns on all 370 stocks and update the portfolios at the first week of January and those of July. Note that we use the first 102 returns as training data for the first estimate. We then use returns recursively following the first estimation. The portfolios we construct are maintained for half a year, and their value-weighted returns are recorded in the last week of December and June. The ex-post standard deviation of the recorded returns on each portfolio is used to represent the performance of the corresponding covariance matrix estimator. 

\subsection{Empirical Results}\label{subsec6.2}
We use seven covariance matrix estimators to create global minimum variance portfolios. First, the sample covariance matrix estimator is used as the simplest method. Next, we consider a set of time-invariant approximate factor estimators: the Fama-French three-factor estimator, the five-factor estimator, the factor estimator, and the PPCA. The most recent characteristics in each training data set is used for the PPCA. Finally, we apply the local PCA and the local PPCA to the covariance matrix estimation, considering the time-varying approximate factor model and the time-varying approximate characteristic-based factor model, respectively. For the local PCA and the local PPCA, we use the Epanechnikov kernel and choose fixed values of $h$ and $C_{NT}$ that minimized ex-post standard deviation. To compare the performance of the covariance matrix estimators over different degree of time variation in true factor loadings as in Section \ref{subsec5.2}, we study the ex-post standard deviation of the portfolios during two periods, January 1994 - December 2000 and January 2006 - December 2012. We anticipate that true factor loadings would be stable during the first period but volatile during the second period.

\setcounter{table}{0}
\begin{table}[h]
\renewcommand{\arraystretch}{1.2}
\captionsetup{font=normalsize}
\centering
\caption{Ex Post Standard Deviation of the Global Minimum Variance Portfolios.}\label{tab1}
\begin{tabularx}{\textwidth}{ X  >{\centering}p{0.25\textwidth}  >{\centering}p{0.25\textwidth}} \hline \hline
& Jan 1994 - Dec 2000 & Jan 2006 - Dec 2012 \tabularnewline \hline
Covariance Matrix Estimator & Std. Dev & Std. Dev \tabularnewline \hline
\multicolumn{3}{c}{Panel A: Time-invariant covariance matrix estimators} \tabularnewline \hline
Sample covariance matrix estimator &45.87  &40.10  \tabularnewline 
Fama-French three factor estimator & 7.59 & 11.07 \tabularnewline
Five-factor estimator & 7.97 & 8.90 \tabularnewline
Factor estimator & 7.88 & 8.15 \tabularnewline 
PPCA estimator & 7.94 & 7.99 \tabularnewline \hline
\multicolumn{3}{c}{Panel B: Time-varying covariance matrix estimators} \tabularnewline \hline
Local PCA estimator & 5.00 & 7.21   \tabularnewline
Local PPCA estimator &5.46 & 6.36 \tabularnewline \hline \hline
\multicolumn{3}{p{0.97\textwidth}}{\small{Note: We construct global minimum variance portfolios based on returns for 370 stocks, using seven covariance matrix estimators. We use weekly data from 370 stocks. collected during two periods, January 1994 - December 2000 and January 2006 - December 2012. We estimate the covariance matrix for the 370 stock returns and then updated the portfolios at the first week of every January and July. The portfolios are maintained for half a year, and their value-weighted returns are recorded in the final week of December and June. The second column reports the ex-post standard deviation of each portfolio during the first period, while the last column reports those recorded during the second period. The ex-post standard deviations are reported as percentage per year.}}
\end{tabularx}
\end{table}
\noindent \indent Table \ref{tab1} displays the ex-post standard deviations of each portfolio. The numbers in the second column report the ex-post standard deviations of the portfolio between January 1994 and December 2000 (the stable period) and the numbers in the last column report those recorded between January 2006 and December 2012 (the crisis period). Panel A of Table 1 contains the time-invariant covariance matrix estimators, which assume that the covariance matrix does not change over time. On the other hand, the estimators displayed in Panel B of the Table 1 allow for time variation in the true covariance matrix. Recall that the ex-post standard deviation of each portfolio is used to measure the performance of the corresponding covariance matrix estimator.

In Table \ref{tab1}, we first observe that the ex-post standard deviation of the portfolio constructed using the sample covariance matrix estimator is much larger than that of the other portfolios. This is consistent with the established belief that sample covariance matrix estimators have poor performance in high dimensional covariance matrix estimation. Table \ref{tab1} also illustrates that the time-varying covariance matrix estimators (Panel B) outperformed the time-invariant covariance matrix estimators (Panel A). Specifically, the ex-post standard deviations of the local PCA and the local PPCA are lower than those of the time-invariant covariance matrix estimators during both periods. This result supports our assumption that both loadings and covariance matrices are time-varying, and changes in both are non-negligible. Finally, Table \ref{tab1} demonstrates that applying characteristics to a covariance matrix estimation is more helpful during a crisis period. The local PPCA outperforms all the other estimators during the crisis period; this is consistent with our simulation result. However, during the stable period, the local PCA outperforms the local PPCA, even though our simulation demonstrates that the local PPCA outperforms the local PCA for all values of $t$. This pattern suggests that it may not be possible to explain all the loadings based on the chosen characteristics. Therefore, if changes in loadings are gradual enough not to require the smoothing effect of the local PPCA, the local PCA may outperform the local PPCA because of the bias caused by the projection of the unexplained part. On the other hand, if loadings fluctuate violently, the benefit of the smoothing projection outweighs the drawbacks posed by the bias and thus the local PPCA can perform better than the local PCA. 

\section{Conclusion}\label{sec7}
This study undertakes the time-varying high dimensional covariance matrix estimation. Working with a time-varying approximate factor model in which the factor loadings, factor covariance matrix, and error covariance matrix change smoothly over time, we propose a covariance matrix estimator. We also introduce another estimator corresponding with a time-varying approximate characteristic-based factor model. Our simulation study demonstrates that characteristics help to estimate factor loadings more precisely in finite samples, making it possible to estimate the covariance matrix more accurately. Moreover, even greater improvement can be achieved when factor loadings are volatile. In the empirical study, we observe that the global minimum variance portfolios constructed by time-varying covariance matrix estimators outperform benchmarks. We also note that the benefit provided by the characteristics increases in crisis periods, which could empirically demonstrate the importance of the simulation result.

\appendix\manuallabel{appen}{Appendix}
\section{: Assumptions and Lemmas in \cite{su2017time}}
We list some assumptions and technical lemmas drawn from \cite{su2017time}. Since we use the local PCA to estimate factors and loadings, the following assumptions are required. Also, we apply the technical lemmas mentioned below to our proofs.\\

\noindent\textbf{Assumption A.1}\manuallabel{a.A}{A.1}\\
\textit{(i) For some $R \times R$ positive definite matrix $\bm{\Sigma}_{\bm{\Lambda}_{r}}$, $\max\limits_{r\leq T}\lVert N^{-1}\bm{\Lambda}_{r}'\bm{\Lambda}_{r}-\bm{\Sigma}_{\bm{\Lambda}_{r}}\rVert=o(1)$ and the eigenvalues of $\bm{\Sigma}_{\bm{\Lambda}_{r}}$ are bounded below from 0 and above from infinity uniformly in $r$.}\\
\textit{(ii) $E(u_{it})=0$, $\max\limits_{i\leq N,t \leq T}{E(u_{it}^{8})}<\infty$, and $\max\limits_{t\leq T}E\lVert \bm{f}_{t}\rVert^{8} < \infty$.}\\ 
\textit{(iii) $\bm{\lambda}_{it}$ are nonrandom such that $\max\limits_{i\leq N,t\leq T}\lVert \bm{\lambda}_{it} \rVert \leq C<\infty$.}\\
\textit{(iv) $\max\limits_{t\leq T}\sum_{s=1}^{T}|Cov(f_{mt}f_{nt},f_{ms}f_{ns})|\leq C$ for $m,n=1, \ldots, R$, where $f_{mt}$ denotes the $m$th element of $\bm{f}_{t}$.}\\
\textit{(v) Define $\psi_{N}=N^{-1}E(\bm{u}_{s}'\bm{u}_{t})$, $\psi_{N,F}=N^{-1}E(\bm{f}_{s}\bm{u}_{s}'\bm{u}_{t})$, and $\psi_{N,FF}=N^{-1}E(\bm{f}_{s}\bm{u}_{s}'\bm{u}_{t}\bm{f}_{t}')$.}\\
\textit{$\max\limits_{t\leq T}\sum_{s=1}^{T}\lVert \psi (s,t)\rVert \leq C$ and $\max\limits_{s\leq T}\sum_{t=1}^{T}\lVert \psi (s,t)\rVert \leq C$ for $\psi(s,t)=\psi_{N}$, $\psi_{N,F}$, and $\psi_{N,FF}$.}\\
\textit{(vi) $\max\limits_{r,t\leq T}E\lvert N^{-1/2} \bm{\Lambda}_{r-1}'\bm{u}_{t}\rvert^{4}\leq C$, $\max\limits_{t,s\leq T}|\zeta_{st}|=O_{p}(\sqrt{\log{T}/N})$, and $\max\limits_{s,t\leq T}E\vert N^{1/2}\zeta_{st} \vert ^{4} \leq C$, where $\zeta_{st}=N^{-1}\{ \bm{u}_{s}'\bm{u}_{t}-E(\bm{u}_{s}'\bm{u}_{t}) \}$.}\\
\textit{(vii) Let $\overline{\omega}_{NT,1}(r)=\sqrt{h/(NT)}\bm{F}^{(r)'}\bm{U}^{(r)}\bm{\Lambda}_{r-1}$ and $\overline{\omega}_{NT,2}(r,t)=\sqrt{h/(NT)} \{ \bm{F}^{(r)}\bm{U}^{(r)}\bm{u}_{t}-E(\bm{F}^{(r)}\bm{U}^{(r)}\bm{u}_{t}) \}$. $\overline{\omega}_{NT,1}(r)=O_{p}(1)$ and $\max\limits_{r,t\leq T}\lVert \overline{\omega}_{NT,2}(r,t) \rVert=O_{p}(\sqrt{\log{T}})$ for each $r$.}\\
\textit{(viii) As $(N,T) \rightarrow \infty,\, h \rightarrow 0,\, Th \rightarrow \infty$, and $Nh \rightarrow\infty$.}\\

Note that $\bm{V}_{NT}^{(r)}$ denotes a $R\times R$ diagonal matrix of the first $R$ largest eigenvalues of $(NT)^{-1}\bm{Y}^{(r)'}\bm{Y}^{(r)}$ in descending order. $\bm{V}_{r}$ is a diagonal matrix consisting of the eigenvalues of $\bm{\Sigma}_{\bm{\Lambda}_{r}}^{1/2}\bm{\Sigma}_{\bm{f}_{r}}\bm{\Sigma}_{\bm{\Lambda}_{r}}^{1/2}$ in descending order, and $\bm{\Upsilon}_{r}$ is the corresponding eigenvector matrix. Let $\bm{Q}_{r}=\bm{V}_{r}^{1/2}\bm{\Upsilon}_{r}\bm{\Sigma}_{\bm{\Lambda}_{r}}^{-1/2}$ and $\bm{H}_{r}=(NT)^{-1}\bm{\Lambda}_{r-1}'\bm{\Lambda}_{r-1}\bm{F}^{(r)'}\widehat{\bm{F}}^{(r)}\bm{V}_{NT}^{(r)\,-1}$.

\begin{flalign*}
&\textbf{Lemma A }(i)\,\bm{V}_{NT}^{(r)}=\bm{V}_{r}+O_{p}(\frac{1}{\sqrt{N}}+\frac{1}{\sqrt{Th}}).\quad (ii)\, \bm{H}_{r}=\bm{Q}_{r}^{-1}+O_{p}(\frac{1}{\sqrt{N}}+\frac{1}{\sqrt{Th}}).&\\
&(iii)\,\frac{1}{NT}\sum_{t=1}^{T}\lVert \bm{\Lambda}_{r-1}'\bm{u}_{t}^{(r)} \rVert^{2}=O_{p}(1).\quad (iv)\,\frac{1}{T^{2}}\sum_{s=1}^{T}\sum_{l=1}^{T}k_{h,sr}k_{h,lr}(\sum_{t=1}^{T}k_{h,tr}\zeta_{st}\zeta_{lt})^{2}=O_{p}(\frac{T^{2}}{N^{2}}).&\\
&(v)\, \frac{1}{N^{3}T^{3}}\sum_{i=1}^{N}\sum_{s=1}^{T}\lVert \sum_{t=1}^{T}E[\bm{u}_{s}^{(r)'}\bm{u}_{t}^{(r)}]u_{it}^{(r)}\rVert^{2}=O_{p}(\frac{1}{(Th)^{2}}).&\\
&(vi)\, \frac{1}{T^{3}}\sum_{i=1}^{N}\sum_{s=1}^{T}\lVert \sum_{t=1}^{T}(\bm{u}_{s}^{(r)'}\bm{u}_{t}^{(r)}-E[\bm{u}_{s}^{(r)'}\bm{u}_{t}^{(r)}])u_{it}^{(r)}\rVert^{2}=O_{p}(1). &\\
&(vii)\,\frac{1}{N^{2}T^{3}}\sum_{s=1}^{T}\lVert\sum_{t=1}^{T}\bm{u}_{s}^{(r)'}\bm{u}_{t}^{(r)}\bm{f}_{t}^{(r)'}\rVert^{2}=O_{p}(\frac{1}{T^{2}h^{2}}+\frac{1}{N}). &\\
&(viii)\, \frac{1}{NT^{2}}\lVert\sum_{t=1}^{T}\sum_{s=1}^{T}\bm{f}_{s}^{(r)}\bm{u}_{s}^{(r)'}\bm{u}_{t}^{(r)}\bm{f}_{t}^{(r)'}\rVert\}=O_{p}(\frac{1}{Th}).&
\end{flalign*}

\section{: Supplementary Appendix}\manuallabelB{suppl}{supplementary appendix}
Supplementary appendix contains all proofs and technical lemmas for this paper. The paper can be downloaded at \url{https://drive.google.com/open?id=1Hhw_2TFqgV3fcPJN0CzuMsK7nE-Hcxr8}. 

\bibliographystyle{model2-names}






\end{document}